\documentstyle[12pt]{article}
\author{Hans - J\"urgen Schmidt}
\title{The classical solutions of two-dimensional gravity}  
\date{}
\begin{document}
\maketitle

\centerline{Universit\"at  Potsdam, Institut f. Mathematik}
\centerline{D-14415 POTSDAM, PF 601553, Am Neuen Palais 10,
Germany}
\centerline{e-mail: hjschmi@rz.uni-potsdam.de}

\bigskip

\centerline{and}

\bigskip

\centerline{Freie Universit\"at Berlin, Institut f. 
Theoretische 
Physik}
\centerline{D-14195 BERLIN, Arnimallee 14, Germany}

\renewcommand{\baselinestretch}{1.2}
\newcommand{\be}{\begin{equation}}
\newcommand{\ee}{\end{equation}}
\begin{abstract}
The solutions of two-dimensional gravity following from
a non--linear Lagrangian  ${\cal L}=f(R) \sqrt{g}$ are
classified, and their symmetry and singularity properties 
are described. Then  a conformal transformation is applied to
rewrite these solutions   as  analogous solutions 
of two-dimensional Einstein-dilaton gravity and vice versa.
\end{abstract}

\noindent 
KEY : Dilaton gravity in 1+1 dimensions, exact solutions,

Birkhoff theorem, conformal transformation

\bigskip

\noindent
Preprint UNIPO-MATH-98-June-8

\noindent 
to appear 1999 in: Gen. Rel. Grav. {\bf 31}

rec. June 8, 98, revised February 21, 99

\bigskip

\section{Introduction}
\setcounter{equation}{0}

The classical solutions of gravity theories in one spatial and
one temporal dimension and their properties have been
discussed recently under the following three points of view:

\medskip

1. As dimensionally reduced higher-dimensional models with
symmetries; the most often used example is the reduction of
3+1-dimenional spherically symmetric space-times to 1+1
dimensions. 

\medskip

2. As model inspired by one of the classes of string or 
superstring theories. 

\medskip

3. As a toy model for the quantization of 3+1-dimensional
gravity, especially for the process of black hole evaporation.

\bigskip

In papers following one of these last two points of view, the
emphasis is mainly on the quantization, and the classical
behaviour is often only  touched. Therefore, many of the
 classical results are hidden in footnotes or appendices to
such papers. 

\bigskip

It is the aim of the present paper to concentrate on 
this classical behaviour independently of an answer to
the question to which of these three points of view 
the results shall apply. We shall start with the following
model:

Let $f(R)$ be any given smooth 
function of the curvature scalar $R$ of a two-dimensional
 (Pseudo--)Riemannian 
manifold $V_2$ with metric $g_{ij}$, ($i,j=1,2$) and let 
$g =  \vert \det g_{ij} \vert $. \footnote{Smooth 
means  $C^{\infty}$, and if $f$ is only smooth
for a certain interval of $R$-values, then we restrict
the discussion to just this interval.} Let
\be
{\cal L} \qquad = \qquad f(R) \sqrt g
\ee
be the Lagrangian, and for compact spaces $V_2$ 
 the corresponding action is
\be  
I  \qquad = \qquad \int_{V_2} \ {\cal L} \ d^2 x
\ee
Usually, this action integral is also written
for noncompact spaces  $V_2$; however, for these
cases, $I$ need not be well-defined.

The Euler--Lagrange equation (equivalently called:
field equation)  descri\-bes the vanishing of 
the variational derivative of the Lagrangian eq. (1) 
with respect to the metric $g_{ij}$. For compact 
spaces $V_2$ this takes place if and only if the action
$I$ eq. (2) has a stationary point there. For 
noncompact spaces $V_2$ the variation is made for suitably
chosen subspaces only; the procedure is done in two
steps as follows: First, let $K \subset V_2$ be any 
compact subset, then a metric $g_{ij}$ is called
$K$-stationary if it makes the $K$-action
$$
I_K  \qquad = \qquad \int_{K} \ {\cal L} \ d^2 x
$$
stationary. Second, $g_{ij}$ is called stationary, if it is 
$K$-stationary for every such compact set $K$.

\bigskip

The trace of the Euler--Lagrange equation reads
\be 
0 \qquad = \qquad G \ R \ - \ f(R) \ + \ \Box G
\ee
where
\be 
G   \qquad = \qquad  \frac{df}{dR}
\ee
and $\Box$ denotes the D'Alembertian. The remaining part of
the field equation is equivalent to the vanishing of the
tracefree part\footnote{If one looks into
the deduction of the field equation one can see: 
This property is a consequence of the fact that the tracefree
part of the Ricci tensor identically vanishes in two
dimensions.} of the tensor $G_{;kl}$, where the semicolon
denotes the covariant derivative. 
\bigskip

The curvature scalar $R$ in $V_2$ is just the double of
the Gaussian curvature, therefore, 
$$
\int_{V_2} \ R \ \sqrt g \  d^2 x
$$
represents a topological invariant.
In fact, it is a multiple of the Euler characteristic 
if $V_2$ is compact and $g_{ij}$ is positive definite.

\bigskip

If $f$ is a linear function of $R$, then $G$ eq. (4)
is a constant and the action $I$ eq. (2)
is simply a linear combination of the volume of $V_2$
and its Euler characteristic.  For this case, the field
equation 
has either no solution, or every $V_2$ represents a solution.
Therefore, we assume in the following that 
\be 
\frac{d^2 \ f}{d \ R^2} \qquad \ne \qquad 0
\ee
\medskip

The multiplication of the action by a non--vanishing 
constant does not alter the set of solutions of the field
equation. Together with the above we have 
now justified to define:

Let $\alpha$ and $\beta$ be two constants with 
$\alpha \ne 0$, then the functions $f(R)$ and
$\alpha \ f(R) \ + \ \beta \ R$ are considered as equivalent.

\bigskip

The remaining part of this 
 Introduction presents short comments to the cited literature:
The classification given below is only a rough one due to
 the fact that the majority of papers contributes to more than
one of the mentioned topics. And, in many cases one should 
have added ``and the references cited there'' to get a
more complete reference list. 

\bigskip

Refs. [1-4] consider 2-dimensional gravity from the point of
view as dimensionally reduced higher-dimensional gravity
models, ref. [1] compares 2- and 3-dimensional theories, 
[2] compares with the spherically symmetric solutions of 
$d$-dimensional Einstein theory ($d>3$), and 
[3] with the analogous case in the 
 $d$-dimensional Einstein--dilaton--Maxwell theory.
 Ref. [4] makes the dimensional reduction from Einstein's 
theory in $d$ dimensions for those $d$-dimensional 
space--times where $n=d-2$ commuting hypersurface--orthogonal 
Killing vectors exist, to metric--dilaton gravity in 2 
dimensions with $n$ scalar fields.  

\medskip

A slightly different point of view to dimensionally reduced
models can be found in [5, 6]; but also in these two papers,
cosmological models are dimensionally reduced to 2-dimensional
dilaton theories. 

\bigskip

Refs. [7-14] are mainly concerned with the higher order
theory in two dimensions. [7,8] deal with the 
discretized version in Regge calculus, [7] with Lagrangian
$R^2$, 
[8] with the more general scale-invariant 
Lagrangian $R^{k+1}$. 

\medskip

Papers [9-14] consider the classical version 
of the theories following from nonlinear Lagrangian $f(R)$,
and the present paper completes the discussion started
in [9], with subsequent papers [10, 12, 14]. In [11], the
Lagrangian $R \ln R$ plays a special role, and in [13],
for $R^{k+1} + \Lambda$ the solutions have been given in
closed form (but not in full generality).

\bigskip

Refs. [15-41] deal with Einstein--dilaton
 gravity in 2 dimensions
mainly from the classical (i.e., non--quantum) point of view.
Due to the equivalence of this theory to the above-mentioned
nonlinear fourth--order theory (see e.g. [12, 34] and the
present paper) many of the results are parallelly developed.
 
\medskip

The CGHS-theory [15] was the starting point for several other 
papers. Further papers on this topic are [16-20]. 
In [18], the global behaviour of the solutions is 
discussed from the ``kink'' point of view: this refers to
space--times with twisted causal structure.

Related papers have themes 
 as follows. In Ref. [21]: wormholes; in refs.[22-29]: black
holes; in refs. [30,31]: collapse behaviour; 
in refs. [32-35] the general solutions have been discussed; 
and a more general discussion about such models can be found
in ref. [36-41]. 

\bigskip

Theories including torsion are discussed in
refs. [42-46], [46] represents a review about this topic.

\bigskip

In refs. [47-60], quantum aspects play the major role, but
in all of these papers, the classical aspects are at least
mentioned as a byproduct. The main topics of them are as
follows: 
Supergravity [47], entropy [48],
gravitational anomaly [49-51], quantization procedure
 [52-55], and evaporation of black holes [56-59]. 
For more details see the review [60].

\bigskip

Refs. [61-68] deal with the differential geometric 
points of view. In [61], the inequivalence of different
 definitions for the stationarity of the action is shown;
 in [62], asymptotic symmetries are considered; 
 in [63], warped products of manifolds and conformal
transformations are used to relate several models into each
other; and
  Killing tensors in 2 dimensions are deduced in [64]. 

\medskip

Ref. [65]  deduced curvature properties and singularity
behaviour of several 1+1-dimensional space--times with one 
Killing vector. It is not directly related to physics,
and no field equations are considered. Nevertheless, 
its results can be  directly applied to several of the
solutions of 2-dimensional gravity. The geodesics and
their completeness has been discussed in refs. 
[13,65-68].  
 
\bigskip

In [69], a canonical transformation from dilaton gravity
 into a free field theory is given. 

\bigskip

The paper is organized as follows: 
Sections 2 till 6 deal only with the fourth--order 
theories according to eqs. (1, 2), Sections 7 till 10 
also with dilaton theories. 

\medskip

In more details: Sct. 2 deals  
with the Birkhoff theorem in 2 dimensions and gives a
coordinate--independent proof of it, see the key
equation (6). Sct. 3 gives a method to integrate the field
equation in Schwarzschild coordinates. The Killing vector
from sct. 2 explicitly serves to simplify the deduction, see
eq. (12). For the general model eq. (1) the complete solution
can be given in closed form, eqs. (10, 14). Sct. 4
concentrates on the scale-invariant case, i.e. $f(R)=R \ln R$
or $f(R)=R^{k+1}$. Sct. 5 enumerates the corresponding
solutions, and Sct. 6 describes their differential geometric
properties. 

\medskip

In Sct. 7 both the transformation from the model eqs. (1, 2)
to dilaton gravity and the corresponding back transformation
are explicitly given. Sct. 8 applies this transformation to
the examples discussed in Sct. 3 and 4, and gives a
typical example of a field redefinition. In 
Sct. 9 a conformal transformation is applied which mediates
between different types of dilaton gravity, and in Sct. 10,
this conformal transformation is applied to the solution given
in Scts. 5 and 6. Section 11 discusses the results.

\section{The Killing vector} 

Let $\varepsilon_{ij}$ be the 
antisymmetric Levi--Civita pseudo--tensor
in $V_2$; it can be defined as follows: In a 
right-handed locally cartesian coordinate system it 
holds: $\varepsilon_{12} = 1$. \footnote{Here, a 
coordinate system is
called locally cartesian, if at  
this point, $g_{ij}$ has diagonal form, and the 
absolute values of its diagonal terms are all equal to 1.
 It should be mentioned that the Levi--Civita pseudo--tensor
is defined for oriented manifolds only. For the other
cases there remains a sign ambiguity; however, in what
follows, this ambiguity does not influence the 
deduction.}  It holds 
$$
\varepsilon_{ij;k} \ =  \ 0 \ .
$$
 Now we define with $G$ from eq. (4) 
\be 
\xi_l \qquad = \qquad \varepsilon_{lm} \ G^{;m}
\ee
The {\it Birkhoff theorem in two dimensions} (see e.g. 
[9,23])
states that 
locally, every solution of the field equation \footnote{
Here ``field equation'' is used in the sense ``vacuum field
equation''; the inclusion of matter would, of course, alter
the result, but this is not topic of the present paper.}
 possesses 
an isometry. 

{\it Proof:}  If $G$ is constant over a whole region 
(that means, $\xi_l = 0$ there)
then because of inequality (5), $R=$ const., i.e.,  
locally, the space is of constant curvature and 
possesses a 3-dimensional isometry group.
  If $\xi_l$ eq. (6) is a non--vanishing but
light--like vector over a whole region 
 then $V_2$  is locally of constant curvature, too.
 So, apart from some singular points and lines where $\xi_l$
vanishes or is light-like, the  vector $\xi_l$ may be assumed
to be a non-vanishing time-like or space-like vector and  
 it suffices to show that it represents a Killing vector. To
this end we calculate
\be 
\xi_{l;k} + \xi_{k;l} = \varepsilon_{lm} G^{;m}_{;k}
+ \varepsilon_{km} G^{;m}_{;l}
\ee
The vanishing of the tracefree part of $G_{;ij}$ is
equivalent to the existence of a scalar $\Phi$ such that
$G_{;ij} = \Phi g_{ij}$. We insert this into the r.h.s. of eq.
(7) and get
$$
= \Phi \varepsilon_{lm} \delta^m_k
+ \Phi \varepsilon_{km} \delta^m_l
= \Phi (\varepsilon_{lk} + \varepsilon_{kl}) = 0
$$
q.e.d.

Further, it holds: 
The field equation if fulfilled iff (= if and only if)
the trace equation (3) is fulfilled and the vector
defined by eq. (6) represents a Killing vector.

{\it Proof:} It remains to show that 
$\xi_{l;k} + \xi_{k;l} = 0$ implies 
the vanishing of the tracefree part of $G_{;ij}$. 
To this end we introduce the inverted Levi--Civita
 pseudo--tensor $\varepsilon^{lm}$ via 
$$
\varepsilon^{lm} \ \varepsilon_{mk} \ = \ \delta^l_k
$$

and get from eq. (6)
\be 
G^{;m} = \varepsilon^{ml} \ \xi_l
\ee
and after applying \ ``$;k$'' \  we 
get the requested identity.
 q.e.d.

\bigskip

As a corollary  from this proof we get 
\be 
\Box \ G \qquad  = \qquad 2 \ \varepsilon^{12} \ \xi_{2;1}
\ee


\section{Schwarzschild coordinates} 

In this section, we consider solutions of the field
equation  in such a region where the Killing vector
$\xi_l$ eq. (6) is a non--vanishing time--like or space--like
vector. Then, locally, we may use Schwarzschild
coordinates
\be 
ds^2=g_{ij}dx^idx^j=\frac{dw^2}{A(w)} \pm A(w) \ dy^2
\ee
The overall change $ds^2 \longrightarrow - ds^2$ 
does not represent an essential change, so we have to deal
with two signatures: the upper sign in eq. (10)
gives the Euclidean, the lower sign gives the Lorentzian 
signature.  
 Here we concentrate on 
 the Euclidian signature case only, but locally, an
imaginary 
transformation $y\to iy$ gives all the corresponding Lorentz
signature solutions, too.

We assume $(w=x^1, \ y = x^2)$ to represent a right--handed
system. Therefore, in the coordinates of metric (10) we
have $\varepsilon_{12} = 1, \  
\varepsilon^{12} = -1$.

From metric (10) we get 
\be 
R \qquad =  \qquad - \  \frac{d^2A}{dw^2},
\ee
The constant curvature cases are already excluded, so we 
have to assume that $A(w)$ is {\it not} a polynomial
 of degree 2 or less. Under these circumstances, metric
(10) has exactly one isometry, a translation into the
$y$--direction reflecting the fact, that $g_{ij}$ does not
depend on $y$,  i.e. $\xi^i = \alpha (0,1)$, where 
$\alpha$ is a non-essential non-vanishing constant. 
We get
\be 
\xi_i \quad = \qquad (0, \ \alpha \ A(w))
\ee
We insert eq. (12) into eq. (8) and 
get $G_{;1} = - \alpha$, 
 i.e., $G$ is a linear but not constant function of  $w$.
By a linear transformation of $w$ which does not alter the
ansatz eq. (10)  we can achieve\footnote{with $\alpha = -1$
this is consistent with eq. (6)}  $G(R)=w$. This equation can
be, at least locally, inverted 
to 
$$
R=\psi(w). 
$$
From eq. (9) we get $\Box G=\frac{dA}{dw}$, and then the
trace (3) of the field equation reads 
\be 
0=w\cdot \psi(w)-f(\psi(w))+\frac{dA}{dw}
\ee
We introduce the integration constant $C$ and get 
\be 
A(w) \qquad = \qquad C \ + \ \int f(\psi(w))-w\cdot \psi(w)dw
\ee
which represents the general solution to the field equation.

\bigskip

\noindent {\it Example:}

Let us take $f(R)=e^R$ and apply the above procedure 
to this Lagrangian. From eq. (4) we get $G=e^R=w>0$, i.e.,
$R=\psi(w)=\ln w$ and $f(\psi(w)) = w$. 
From eq. (14)  we get 
$$
A(w) \ = \ C-\frac{w^2}{2}\ln w + \frac{3w^2}{4}
$$
via $\frac{dA}{dw}=w-w \ln w$, so $A(w)$ has a local 
extremum at $w=e$, i.e. $R=1$. It turns out to be a maximum.
 In dependence on the value of $C$, different 
types of solutions can be constructed. \footnote{This is a 
quite general property of these models: Adding a constant $C$ 
to $A$ eqs. (10,14), then these two space-times are no more
isometric in general, but, according to eq. (11), they have
the same curvature. This is also interesting from the purely
differential geometric point of view: under which
circumstances the curvature uniquely determines the metric,
cf. e.g. [65]}
 In contrast to the 
models to be discussed in the next sections, 
$R=0$ does not play a special role here.

\section{Scale-invariant fourth-order gravity} 

A field equation is scale invariant if the following holds: If
a metric $g_{ij}$ is a solution and $c\neq 0$ a constant, then
the homothetically equivalent metric $c^2 g_{ij}$ is a
solution, too. It holds, cf. [12]:  
$ f(R) =R \ln \vert R \vert $ and $\vert R \vert^{k+1}$
represent the only cases that lead to 
scale-invariant field equations. 
We define for an arbitrary real $k\neq -1$
\be 
f(R)=\left\{ 
\begin{array}{ll} 
 R \ln|R|-R & k=0  \\
 \frac{1}{1+k}R|R|^k & k\neq 0  \\ 
\end{array} 
\right.
\ee
This covers all cases with a scale-invariant
field equation. 
In most cases, $R=0$ represents a singular point of the field
equation, and for those cases we 
restrict to the range $R\neq 0$ and allow $R \to 0$ in the
solutions only afterwards. From eq. (15) we get with (4)  
\be 
G(R)=\frac{df}{dR}=\left\{ 
\begin{array}{ll} 
\ln|R| & k=0  \\
|R|^k & k\neq 0  \\ 
\end{array} 
\right.
\ee
The trace (3) of the field equation reads
\be 
0=R+\Box\:(\ln|R|),\; k=0
\ee
\be 
0=R+\left(1+\frac{1}{k}\right)|R|^{-k} \: \Box\: (|R|^k),\;
k\neq
0
\ee
Instead of explicity solving equation (14) it proves
useful to insert (10) into (17), (18), and we get
\be 
\frac{dA}{dw}\cdot \frac{d^3A}{dw^3}=\frac{1}{k+1}
\left(\frac{d^2A}{dw^2}\right)^2
\ee 
This equation is valid for all values $k$ and for both signs
of $R$.
Even the limit $k\to \pm \infty$ makes sense: In this limit we
get 
$d^3A/dw^3=0$ which is equivalent to the requirement
that the  2-space is of constant curvature.

It is remarkable that the case $k=0$ from equation (15)
is now smoothly incorporated in equation (19).

The limit $k\to -1$ leads to $d^2A/dw^2=0$, i.e. flat
space.

\section{Solutions of scale-invariant gravity} 

Equation (19) has a 3-parameter set of solutions. The
integration
constants are $C$, $D$, $E$, and we get
\be 
A(w)=C+D\cdot \ln|w-E|,\; k=-\frac{1}{2}
\ee
\be 
A(w)=C+D\cdot e^{E\cdot w},\; k=0
\ee
\be 
A(w)=C+D\cdot |w-E|^{2+\frac{1}{k}}\; \; \mbox{else}
\ee
Now, the two signatures of the metric have to be 
distinguished.

\subsection{The Euclidean signature} 

The following examples of solutions are especially
interesting:  
\be 
k=-\frac{1}{2}: \quad 
ds^2=\frac{dw^2}{\ln w}+\ln w \: dy^2,\; w>1,\;
R=\frac{1}{w^2} 
\ee
\be 
k=0: \quad ds^2 = \frac{dw^2}{e^w}+e^wdy^2,\; R=-e^w
\ee
and for the other $k$-values: 
\be 
ds^2=\frac{dw^2}{w^{2+\frac{1}{k}}}+w^{2+\frac{1}{k}}dy^2,\;
w>0, \ 
R=-w^{\frac{1}{k}}\left(1+\frac{1}{k}\right)
\left(2+\frac{1}{k}\right).
\ee

Equations (24), (25) can together be written in conformal
coordinates as follows 
\be 
ds^2=x^{\frac{1}{k+1}}\cdot \frac{dx^2+dy^2}{x^2}, \qquad x>0
\ee
In synchronized coordinates this reads 
\be 
ds^2=dr^2+r^{-4k-2}dy^2, \qquad r > 0
\ee
However, in both cases, the limit $k \longrightarrow -1/2$ 
does not lead to the solution eq. (23). A direct calculation
shows the following: eq. (23) can be written in conformal or
synchronized coordinates only by the use of logarithmic
integrals.

\bigskip

With these examples eqs. (23 - 25), however, the set of
solutions is not exhaustet. Let us complete their enumeration
as follows: $D=0$ gives flat space only, so we exclude this
case. A translation of $w$ can be used to get $E=0$
in (20, 22), and to get $C/D \in \{-1,0,1\}$ in (21).
Then $w$ can be multiplied by a suitable factor to get
$C/D \in \{-1,0,1\}$ in (20, 22) and to get $E=1$ in (21).

Finally, we apply the following point of view: homothetically 
equivalent metrics are considered as equivalent. So, $A$ can
be multiplied by any non-vanishing constant (and $y$ has to
be divided by the same constant). So, we always get 
$D \in  \{-1,1\}$, and we get a finite list of solutions as
follows: 

\bigskip

\noindent For $k = - 1/2$: With $C=0$ we get (23) and its 
counterpart 
\be 
ds^2=\frac{dw^2}{\vert \ln w\vert}+\vert\ln w \vert\ dy^2,\;
0<w<1,\;
R= - \frac{1}{w^2} 
\ee
With $C \in \{-1,1\}$ we get 4 further metrics, however: all
of them are isometric to (23) or (28) which can be seen 
by multiplying or dividing $w$ by $e$.

\bigskip

\noindent For $k= 0$: With $C=1$ we get
\be 
 ds^2 = \frac{dw^2}{1 \pm e^w}+ (1 \pm e^w) 
dy^2, \ (w<0 \mbox{ for lower sign}), \ R= \mp e^w, 
\ee
With $C=-1$ we get
\be 
 ds^2 = \frac{dw^2}{e^w-1}+(e^w-1)dy^2, \ w>0, \ R=-e^w,
\ee
i.e., together with metric (24) we have 4 cases.

\bigskip

\noindent For all other values $k \ne - 1$: With $C=1$ we get 
\be 
ds^2=\frac{dw^2}{1 \pm w^{2+\frac{1}{k}}}
+ \left( 1 \pm w^{2+\frac{1}{k}} \right) dy^2,
\ee
and with $C=-1$ 
\be 
ds^2=\frac{dw^2}{w^{2+\frac{1}{k}} - 1}
+ \left( w^{2+\frac{1}{k}} - 1 \right) dy^2,
\ee
i.e., for every value $k$ we have together with metric (25)
again 4 cases. The range of $w$ has to be chosen
within the interval $w>0$ such that $A>0$. $R$ is, up to the
sign, always the same as in eq. (25). 

If $1/k$ represents an odd natural number, then the two signs
of eq. (31) give solutions which can be smoothly pasted
together at the line $w=0$ (after redefinition 
$w \longrightarrow -w$ in one of the parts), so the number
of solutions is reduced by one for these cases.

\subsection{The Lorentzian signature} 

As already mentioned, the corresponding Lorentzian
solutions can be obtained from the Euclidean ones by
an imaginary rotation of the coordinate $y$. However, after
this rotation, we have more cases where solutions can be
pasted together. This is connected with regular zeroes 
of the function $A$, where the character of the Killing vector
$\xi_k$ changes from time-like to space-like. The line where
this happens is called horizon. Let us summarize these
solutions:

\bigskip 

\be 
k=-\frac{1}{2}: \quad 
ds^2=\frac{dw^2}{\ln w} \quad - \quad \ln w \ dy^2, \qquad w>0
\ee
represents the Wick rotated solutions (23) and (28) pasted
together at $w=1$.

\bigskip

\be 
k=0: \quad ds^2 = \frac{dw^2}{e^w} \quad - 
\quad e^wdy^2, 
\ee

is taken from (24), and 
\be 
k=0: \quad ds^2 = \frac{dw^2}{1 \pm e^w} \quad - 
\quad (1 \pm e^w) dy^2, 
\ee
puts together solutions (29, 30). 

\bigskip

For the other values $k \ne -1$ we get: 
\be 
ds^2=\frac{dw^2}{w^{2+\frac{1}{k}}}
 \quad - \quad w^{2+\frac{1}{k}}dy^2,
\ee
from (25) and (31, 32) can be pasted together as follows: 
\be 
ds^2=\frac{dw^2}{1 \pm w^{2+\frac{1}{k}}}
\quad - \quad  \left( 1 \pm w^{2+\frac{1}{k}} \right) dy^2.
\ee
The remark from the end of subsection 5.1. applies to this
solution, too. 

\bigskip

Finally, it should be mentioned, that we considered 
solutions being related by the transformation 
$ds^2 \longrightarrow - ds^2$ as equivalent ones. 
 (This transformation  does not change the
signature of the solutions, but it changes the  
sign of the curvature scalar.) 

\section{Properties of these solutions} 

In this section, we discuss the properties of the 
solutions found in sct. 5.  

\subsection{The curvature invariants} 

First of all, let us calculate the curvature invariants
using metric (10) and $R$ from eq. (11). We get
\be 
R_{;i} \ R^{;i} \quad = \quad A \  \left( \frac{d^3 A}{dw^3}
\right) ^2 
\ee
and
\be 
\Box \ R \quad = \quad - A \ \frac{d^4 A}{dw^4}
 \ - \ \frac{d A}{dw} \ \frac{d^3 A}{dw^3} .
\ee
By use of eq. (19) we eliminate $d^3 A/dw^3$
and get
\be 
R_{;i} \ R^{;i} \quad = \quad \frac{A}{(dA/dw)^2} \ \cdot \
\frac{R^4}{(k+1)^2}
\ee
Applying $\frac{d}{dw}$ to eq. (19) we can also 
eliminate 
$d^4 A/dw^4$ and get by use of eq. (40)
\be 
R \ \Box R \quad = \quad - \ \frac{R^3}{k+1}
\ - \ (k-1) \ R_{;i} \ R^{;i}
\ee
Neither $A$ nor $dA/dw$ have an invariant meaning because
they can be changed by a coordinate transformation
$w \to \alpha w$. However, as can be seen from eq. (40), 
for $R \ne 0$ the quotient $A/(dA/dw)^2 $ 
possesses an invariant meaning after $k$ has been fixed. 

\bigskip

For the higher order curvature invariants we get a
result analogous to eq. (41): After $k$ has been fixed,
all of them can be expressed as a function of $R$
and $R_{;i} \ R^{;i}$. For $k \ne 1$, we can also say: 
Every invariant can be expressed by use of $R$ and $\Box R$. 
 So, within these models we get: If one of the curvature
 invariants diverges, then already $R$ diverges. 
In other words: $R$ alone is sufficient to decide 
whether a curvature singularity exists.

\subsection{Selfsimilarity of solutions} 

A solution $ds^2$ of the field equation 
 is called selfsimilar, if for every constant $\alpha >0$
the homothetically equivalent metric $\alpha ds^2$ 
is isometric to $ds^2$. It holds: A space of constant
curvature is selfsimilar iff $R=0$. 
 From the solutions of sct. 5 the
following ones are selfsimilar: (24), (25), consequently also
(26), (27), (34) and (36). 
The remaining ones -- which include all solutions with a
horizon --  are not selfsimilar: (23), (28 - 33),  
 (35) and (37).

\subsection{Geodesics} 

To get a better knowledge about these solutions it
proves useful to calculate their geodesics. This is
necessary, because in many cases, a coordinate tends
to infinity without describing an infinite distance.
 For the Lorentzian case, one has also to distinuish
between completeness of the 3 types of geodesics, the
result may be different, cf. e.g. [13,44,66-68].

\subsubsection{The case $k=-1/2$}
Metric (23) represents a geodesically complete surface 
of infinite surface area and topology $E^2$ of the Euclidean
plane if one point $[w=1]$ is added as symmetry center. 
This can be seen as follows: Starting from an arbitrary point,
 a geodesic with finite natural parameter ends always at a
finite value of the coordinate $w$, so  $w \longrightarrow
\infty$   makes no problem. 

To analyze the neighbourhood of $w=1$ we introduce new
coordinates $z=2\sqrt{w-1}$ and a $2\pi$-cyclic 
coordinate $\phi = y/2$. The eq. (23) reads
$$
ds^2 = \frac{dz^2}{1 - z^2/8 + \Sigma} + z^2(1 - z^2/8 +
\Sigma)d\phi^2
$$
where 
$$
\Sigma = \sum_{n=2}^{\infty} \frac{(-1)^n}{n+1}
 \cdot \frac{z^{2n}}{4^n}
$$
which is regular as $z\to 0$ and rotationally symmetric
with $[z=0]$ as center of symmetry.

\bigskip

Metric (28) can be analyzed quite similarily: If we put
$v=1/w$ then we get
$$
ds^2=\frac{dv^2}{v^4 \ln v} + \ln v dy^2, \qquad v>1, \qquad
 R = - v^2
$$
and the behaviour for $v\to 1$ is the same as for $w\to 1$ in
the above example from eq. (23). But for $v \to \infty$ 
we have a curvature singularity in a finite invariant distance

which can be seen by calculating the length of the geodesic
 $[y=0]$. The surface has topology $E^2$ and surface area
$4\pi$. 

\bigskip

The Lorentzian signature solution (33) can be analyzed by
introducing Bondi coordinates, i.e., for $w>1$ we define
$$
u \qquad = \qquad y \ - \ \int \  \frac{dw}{\ln w}
$$
instead of $y$, and then we apply an analytic continuation
to the whole interval $[w>0]$ afterwards. We get 
(cf. the remark at the end of sct. 5.2.)
$$
ds^2 \ =  \ \ln w \ du^2 \ + \  2 du \ dw, \qquad 
R = \frac{1}{w^2}$$
The line $[w=1]$ represents a regular horizon; 
 $[w=0]$  is a curvature singularity, it will be reached after
finite invariant length of a geodesic, so it represents a true

 curvature singularity; and for $w \to \infty$ the space is
asymptotically uncurved but not asymptotically flat.

\subsubsection{The Euclidean cases $k \ne -1/2$, $C=0$
 }

The cases with $C=0$, where 
 $k\ne - 1/2$ and $k\ne - 1$,  can be
analyzed in synchronized coordinates. For 
$$
ds^2 = dr^2 + a^2 (r) dy^2
$$
we have
$$
R \qquad = \qquad  - \quad \frac{2}{a} \cdot
\frac{d^2 a}{dr^2}
$$
For metric (27) we have $a(r)=r^{-2k-1}$, i.e.
$$
R \qquad = \qquad  - \quad \frac{4}{r^2} (k+1)(2k+1)
$$
leading to a true curvature singularity as
$r \to 0$ which can be reached at finite invariant geodesic 
distance. For $r \to \infty$ the space is asymptotically
uncurved.

\subsubsection{The Lorentzian cases $k \ne -1/2$, $C=0$
 }

These cases are described by eq. (36), but the properties can
easier be analyzed by writing it in synchronized coordinates,
which leads to a Wick-rotated form of metric (27).
The curvature behaves exactly as in subsection 6.3.2.

\subsubsection{The Euclidean cases $k \ne -1/2$, $C \ne
0$}

For metric (29-32) the properties are as follows:

\medskip

\noindent Metric (29) (i.e. $k=0$, $C=1$):

\medskip

 upper sign: $w\to \infty$ represents a true 
curvature singularity in finite invariant distance. 

\medskip

 lower sign: $w\to - \infty$ gives an asymptotically flat
surface, whereas the limit $w\to 0$ gives a regular surface
if one point $[w=0]$ will be attached as center of symmetry
 and the coordinate $y$ will be considered to be a cyclic one.
This is analogous to the case with $k=-1/2$ discussed above,
the only difference is that we have  now $(n+1)!$
instead of $n+1$ in the denominator of $\Sigma$: 
$$
ds^2 = \frac{dz^2}{1 - z^2/8 + \Sigma} + z^2(1 - z^2/8 +
\Sigma)d\phi^2
$$
where 
$$
\Sigma = \sum_{n=2}^{\infty} \frac{(-1)^n}{(n+1)!}
 \cdot \frac{z^{2n}}{4^n}
$$
which is regular as $z\to 0$ and rotationally symmetric
with $[z=0]$ as center of symmetry.

\medskip

\noindent Metric (30) (i.e. $k=0$, $C=-1$):

\medskip

has a true curvature singularity as
$w \to \infty$ which can be reached at finite geodesic
distance. The behaviour near $w=0$ is similar as for
eq. (29).

\medskip

\noindent Metrics (31) and (32) (i.e. $k\ne 0$, $C \ne 1$):

\medskip

They have the curvature scalar
$$
R \qquad = \qquad \pm \ w^{1/k} \ (2+\frac{1}{k})
\ (1+\frac{1}{k})
$$
For $k>0$ we get: $R$ diverges as $w\to \infty$, and 
this will be reached at finite geodesic distance, so we have a
true curvature singularity. This applies to (31), upper sign,
and to (32). For (31), lower sign, we have to restrict to the
interval $w<1$. To get a regular behaviour there, we have 
again to attach one additional point $[w=1]$ as center of
symmetry, and to make $y$ a cyclic coordinate. 
For these cases, the solution is regular as $w \to 0$. 
The analytic continuation to negative values $w$ is possible
 if $1/k$ is an integer. Let us present two typical examples:

\medskip

\noindent $k= 1$, then metric (31), lower sign reads 
$$
ds^2=\frac{dw^2}{1 - w^{3}}
+ \left( 1 - w^{3} \right) dy^2, \qquad w < 1
$$
which has a singularity at finite distance 
as $w \to - \infty$. The other cases with odd $1/k$ are
similar.

\medskip

\noindent Even $1/k$, $k= \frac{1}{2(n-1)}$ with an integer
 $n \ge 2$,
 then metric (31), lower sign reads 
$$
ds^2=\frac{dw^2}{1 - w^{2n}}
+ \left( 1 - w^{2n} \right) dy^2, \qquad -1<w<1
$$
This is a metric with mirror symmetry at $w=1$, so we
have to attach two points, $w=1$ and   $w=-1$, as the two
centers of the rotational symmetry. This represents a regular
 solution with finite volume and spherical topology
 $S^2$. The coordinate $y$ has to be cyclic with period
$2\pi / n $ to ensure local regularity at $w=1$, therefore 
the total volume equals $4\pi/n=\frac{8\pi k}{2k+1}$.
As a test one can calculate that really $\int R 
\sqrt g d^2x = 8\pi$ as requested from the Gauss-Bonnet
theorem. (The case $n=1$ would give the standard $S^2$ of
constant curvature.) For these values $k$, the field equation
has a singular point at $R=0$; the regular solution 
given here has always $R \ge 0$, and this singular point is
only touched at the line $[w=0]$. 

\medskip

If  $1/k$ is not an integer, then no smooth continuation
to negative values $w$ is possible.

\bigskip

For $k<0$ it holds: $R$ diverges as $w \to 0$ which can be
reached after finite invariant distance.
 At $R=0$, there is a singular point of the 
field equation.

\subsubsection{The Lorentzian cases $k \ne -1/2$, $C \ne
0$}

These cases are described by eq. (337). With the upper sign,
the analysis is exactly as in sct. 6.3.4., for the lower sign
we have additionally to analyze the horizons at $w=1$. Let
again $2+\frac{1}{k}=2n$, then 
$$
ds^2 = \frac{dw^2}{1-w^{2n}} -  (1-w^{2n})dy^2
$$
goes over to 
$$
ds^2 \ = \ (1-w^{2n}) \ du^2 \ + \ 2 \ du \ dv
$$
via Bondi coordinates
$$
u = y - \int\frac{dw}{1-w^{2n}}
$$
which proves regularity via crossing the horizon at $w=1$ for
positive integers $n$.

\section{Transformations relating to dilaton gravity} 

Now we give a relation of fourth-order gravity, section 1
to 6, to dilaton gravity. This relation is possible for
 those regions where $G \ne  0$, cf. eqs. (1,4).
 Due to inequality (5)
 the equation $G=0$ can be fulfilled at singular lines
 only. Therefore, the transformation to be deduced 
below will be valid ``almost everywhere''.

\subsection{From fourth--order to dilaton gravity}

Without loss of generality let $G>0$, otherwise 
we simply change $f(R)$ to $-f(R)$, cf. eq. (4). 
 Then we define $\varphi$ by
$$
e^{-2\varphi}=G(R) \ .
$$
 We invert this relation
(which can locally be done because of inequality (5)) to
$R=R(\varphi)$ and define
\be 
V(\varphi)=e^{-2\varphi} R(\varphi)-f(R(\varphi))
\ee
Then the  Lagrangian eq. (1) can be written as 
\be 
L(\varphi,g_{ij})= [ e^{-2\varphi} R-V(\varphi) ] \sqrt g
\ee
Now we forget for a moment how we deduced eq. (43) 
and take it as given Lagrangian. 
The variation of $L$ eq. (43) with respect to
$\varphi$ gives
\be 
0=2e^{-2\varphi}R+\frac{dV}{d\varphi}
\ee
The variation of this $L$ with respect to
$g_{ij}$ has 
the trace 
\be 
0=V(\varphi)+\Box (e^{-2\varphi})
\ee
and the traceless part of $\left(e^{-2\varphi}\right)_{;lm}$
has to vanish.

This is the transformation from fourth-order gravity to
dilaton gravity, which can already be found in several
of the cited papers, e.g. [11] \footnote{This paper 
by Solodukhin is the published version of the preprint from
1994 cited in ref. [12].}

\subsection{From dilaton gravity to fourth--order}

To 
go the other direction -- which seems not to be worked
 out so explicitly up to now -- 
let us start from dilaton gravity
eq. (43) and calculate by eq. (44) 
\be 
R \qquad = \qquad 
-\frac{1}{2}\: e^{2\varphi}\: \frac{dV}{d\varphi}
\ee
Now we have to distinguish two cases: If this $R$ is constant,
i.e., eq. (43)
 reads 
$$
L=e^{-2\varphi}(R-\Lambda) \sqrt g
$$
 which is the Jackiw--Teitelboim
theory, then the transformation to fourth-order gravity 
is impossible. In
all other cases, we can locally invert eq. (46) to
$\varphi=\varphi(R)$,
and then we insert this into eq. (43) and get
\be 
{\cal L}=f(R) \sqrt g =L(\varphi(R),g_{ij})
\ee
which is the equivalent fourth-order theory eq. (1).

\bigskip

This transformation transforms not only the actions into
each other, but also the solutions of the field
equation -- unless they belong to the mentioned singular 
exceptions --  are transformed into each other.  

\bigskip

{\it Example:} Let us start from dilaton gravity eq. (43)
  with a potential \footnote{The factors ``2'' are
 inserted for convenience only, and we discuss the
 range $0< \exp(-2\varphi) < \pi/4$, 
i.e. $0<R<1$; 
putting $e^{-2 \varphi} = \Phi$, as will be done in
section 8.3., the potential is just $\sin \Phi $.}
$$
V(\varphi) \ = \ \frac{1}{2} \ \sin (2 e^{-2 \varphi})
$$
From eq. (46) we get $R =  \cos (2e^{-2 \varphi})$, i.e.
$$
\varphi \ = \ - \frac{1}{2} \ln(\frac{1}{2} \arccos R)
$$
leading via eq. (47) to
$$
f(R) \qquad = \qquad \frac{R}{2} \ \arccos R \ - \
 \frac{1}{2} \ \sqrt{1 - R^2}
$$

\section{Examples for  dilaton gravity}

Before we turn to the really interesting cases, we try to
elucidate the procedure by applying it to the example
$f(R) = e^R$ given at the end of sct. 3: On gets
$$
G = e^R = e^{-2\varphi}, \mbox{ i.e.    } R = -2\varphi
$$
With eq. (42) we get $V(\varphi )=e^{-2\varphi}(-1
-2\varphi)$ and then eq. (43) reads
$$
L(\varphi,g_{ij})=  e^{-2\varphi} [ R +1 +2\varphi ] \sqrt g.
$$

\subsection{The scale-invariant case} 

Let us start from the fourth-order theory defined
by eqs. (1, 15) with $k \ne -1$.
 We get from $\varphi = - \frac{1}{2} \ln G$ and eq. (16) 
\be 
R(\varphi)=\left\{ 
\begin{array}{ll} 
 \pm \exp \: (e^{-2\varphi}) & k=0  \\
\pm \, e^{ -2\varphi/k} & k\neq 0  \\ 
\end{array} 
\right.
\ee
where the upper sign corresponds to the case $R>0$, and
for $k=0$ we restrict to the range $\vert R \vert >1$.

Application of equation (42) gives 
\be 
V(\varphi)=\left\{ 
\begin{array}{ll} 
\pm \exp\: (e^{-2\varphi}) & k=0  \\
\pm \,\frac{k}{1+k} \exp \:(-2\varphi
\left(1+\frac{1}{k}\right))
& k\neq 0  \\ 
\end{array} 
\right.
\ee
By a suitable translation of $\varphi$, the factor
$\frac{k}{k+1}$ can be 
made vanish, and so we get from eqs. (43) and (49) 
\be 
L(\varphi,g_{ij})=
[e^{-2\varphi}R \mp \exp \:(e^{-2\varphi}) ] \sqrt g,\; k=0
\ee
and
\be 
L(\varphi,g_{ij})=e^{-2\varphi}\left(R\mp
e^{-2\varphi/k}\right) \sqrt g, \; k \ne 0
\ee
This is the well-known dilaton gravity in an exponential
potential, whereas 
equation (50) is the tree-level string action.

\subsection{Addition of a divergence} 

Let us now look what happens if we apply the transformations
mentioned at the end of sct. 1  before
the conformal transformation of sct. 7 is carried out: 
 The factor $\alpha$ gives nothing but an irrelevant 
translation of $\varphi$. However, $\beta$ will
influence the potential as follows: 
We get $G+\beta$ instead of $G$ and
$$
-\frac{1}{2} \ln\: (e^{-2\varphi}+\beta)
$$
instead of $\varphi$
which represents the transformation called ``new 
non-conformal extra symmetry''
in J. Crux et al., Phys. Lett. B {\bf 402} (1997) 270.

\subsection{A field redefinition for $\varphi$} 

Sometimes, the dilaton is written as $\Phi=e^{-2\varphi}$;
this represents only a field redefinition, so all other 
properties remain unchanged.
Instead of (43)  we get
$$
\hat{L}(\Phi,g_{ij}) \quad = \quad [
\Phi R-\hat{V}(\Phi) ] \sqrt g
$$
where $\hat{V}(\Phi)=V(\varphi)$ at $\Phi=e^{-2\varphi}$.
Variation of this Lagrangian 
with respect to $\Phi$ gives 
$$
R=\frac{d\hat{V}}{d\Phi}
$$
 being equivalent to (44). From the variation with respect to
the metric we get that  the traceless part of $\Phi_{;lm}$ 
has to vanish, and its trace eq. (45) now
simply reads 
$$
0=\hat{V}(\Phi)+\Box\Phi
$$
 so, from eq. (50) we get the $k=0$-result
$$
\hat{L}(\Phi,g_{ij})=[\Phi R \mp e^{\Phi}]\sqrt g,\; \Phi=\ln
|R|
$$
and from eq. (51) the remaining cases
$$
\hat{L}(\Phi,g_{ij})
=\Phi( R \mp \Phi^{1/k})\sqrt g,\; \Phi=|R|^k
$$
Again, $k\to \pm \infty$ gives the Jackiw-Teitelboim theory.

\bigskip

The other example, $f(R)=e^R$, leads to 
$$
\hat{L}(\Phi,g_{ij})
=\Phi( R +1 - \ln \Phi)\sqrt g.
$$

\section{A conformal transformation} 

The theory defined by the Lagrangian (43) is often rewritten
in a conformally related metric  
\be 
\tilde{g}_{ij}=e^{-2\varphi}g_{ij}
\ee
One should observe that the conformal factor uniquely depends
 on the curvature scalar $R$ of the metric $g_{ij}$ because
of $e^{-2\varphi} = G(R)$, cf. sct. 7. This
 conformal relation is globally defined. \footnote{Therefore,
this conformal relation is not only  different but also
different in character to the conformal relation following
from the property that all two-spaces
are locally conformally flat -- because in general, the 
conformal factor for the latter is not globally defined.}

\bigskip

We use the abbreviation 
$(\tilde{\nabla}\varphi)^2=\tilde{g}^{lm}\varphi_{,l}\varphi_{
,m}$ and get from 
equation eq. (43)  via the conditions 
$L =\tilde{L} $, $\sqrt{\tilde g} = e^{-2\varphi}\sqrt g$
 and 
\be
R=e^{-2\varphi}[\tilde{R}+4(\tilde{\nabla}\varphi)^2]
\ee
which follows from (52) now 
\be
\tilde{L}=\left(
e^{-2\varphi}[\tilde{R}+4(\tilde{\nabla}\varphi)^2]-
e^{2\varphi}\:
V(\varphi) \right) \sqrt{\tilde{g}}
\ee
The examples discussed above are transformed to:
$$
 k=0   : \qquad
\tilde{L}= \left(
e^{-2\varphi}[\tilde{R}+4(\tilde{\nabla}\varphi)^2]
\mp \exp \: (2\varphi+e^{-2\varphi
}) \right) \sqrt{\tilde{g}}
$$
and
$$ k\neq 0  : \qquad
\tilde{L}= \left(
e^{-2\varphi}[\tilde{R}+4(\tilde{\nabla}\varphi)^2]
\mp \exp \: (-2\varphi/k) \right) \sqrt{\tilde{g}}
$$
The case 
$k=0$ corresponds to Liouville gravity, and $k=1$, which has
$f(R)=\frac{1}{2}R^2$, to the CQHS-model [15].

\bigskip

To transform the example from the end of sct. 3 
 ${\cal L}=e^R \sqrt g$ we get 
$$
\tilde{L}= \left(
e^{-2\varphi}[\tilde{R}+4(\tilde{\nabla}\varphi)^2]+
2\varphi+1 \right) \sqrt{\tilde{g}}
$$
i.e., simply a linear potential in this conformal picture.

\section{Conformal transformation of  solutions of
scale-invariant gravity}

In this section, we transform some of the solutions of
sections 5 and 6  to
the form $d\tilde{s}^2$ according to section 9.
$$
d\tilde{s}^2=e^{-2\varphi} ds^2, \;
e^{-2\varphi}=G(R)=\left\{ 
\begin{array}{ll} 
\ln|R| \qquad k=0  \\
|R|^k \qquad k\neq 0  \\ 
\end{array}
\right. 
$$
according to equation (16). From eq. (23) we get
 (always up to irrelevant constant factors)
$$
d\tilde{s}^2=\frac{w\: dw^2}{\ln w} + w \ln w dy^2,\;
\varphi=\frac{1}{2} \ln w,\; 
k=-\frac{1}{2}
$$
from eq. (24) we get
$$
d\tilde{s}^2=\frac{w\: dw^2}{e^w}+w e^w dy^2,\; k=0 
$$
and from eq. (25) we get
$$
d\tilde{s}^2=
\frac{dw^2}{w^{1+\frac{1}{k}}}+w^{3+\frac{1}{k}}\:dy^2
$$
for the other values $k \ne -1$.

\medskip

The black hole solution given at the end of sct. 6.3.1. has
with $k=-1/2$ now $e^{-2\varphi} = \vert R \vert^{-1/2}=w$,
i.e.
$$
d\tilde{s}^2 \ = \ w \ \ln  w \ 
 du^2 \ + \ 2w \ du \ dw, \qquad
\varphi = - \frac{1}{2} \ \ln w
$$
keeping the regularity at the horizon $w=1$. 

\medskip

The regular Euclidean solution given at the end of sct. 6.3.4.
has ($k=1/2$)  $R= 12 w^2$, hence
$e^{-2\varphi} = \sqrt R = 2 \sqrt 3 \vert w \vert$, so the 
conformal transformation breaks down at $w=0$, just that line,
where the fourth--order field
 equation had its singular points. 

\medskip

In the metrics of this section, $w$ is no more a coordinate
giving the metric in Schwarzschild form, however, a coordinate
change $\tilde w = \tilde w(w)$ can simply be calculated to
get that form.

\section{Discussion}

Let us  show some  unexpected relations
of space-times discussed in this paper to
models discussed from other points of view: 

\bigskip

1. From the example at the end of sct. 7 (i.e., that one which
has in the dilaton version  simply the 
potential $ \ \sin \Phi$ )
 we can find the
solution with the method of sct. 3 as follows: We get
$G= \frac{1}{2} \arccos R$, i.e., $R=\cos(2w)$, and by 
eq. (14) $A(w) = C + \cos(2w)$. For the case $C=0$, the
metric reads
$$
ds^2 \qquad = \qquad 
 \frac{dw^2}{\cos(2w)} - \cos(2w) dy^2
$$
Replacing $w$ by $z$ according to $2w=\arctan \ \sinh (2z)$
we get
$$
ds^2 \qquad = \qquad \frac{dy^2 - dz^2}{\cosh (2z)}
$$
In a next coordinate transformation we change from the
hyperbolic coordinates $y$, $z$ ($z$ is the parameter for the
hyperbolic rotation), to conformal cartesian coordiantes via
$$
t = e^y \ \cosh z \ , \qquad x = e^y \ \sinh z
$$
and get the metric extensively discussed in ref. 18
from a totally different origin: 
$$
ds^2 \qquad = \qquad \frac{dt^2-dx^2}{t^2 \ + \ x^2} \ . 
$$

\bigskip

2. The metric (35), here deduced as  solution for
the Lagrangian $R \ln R$, has  several 
seemingly unrelated origins: from $c=1$ Liouville
 gravity, from non-critical string theory, from a
 bosonic sigma model,  
and from
$k=9/4$ gauged SO(2,1)/SO(1,1) WZW model see e.g. refs.
[22,24,34] for 
further details. 

\bigskip

3. Finally, one should note that also
2-dimensional gravity with torsion is equivalent to
special types of generalized 2-dimensional dilaton
gravity, cf. [44,45].

\section*{Acknowledgement}

I thank Claudia Bernutat for independently checking 
 the essential calculations and Miguel Sanchez  
 for useful comments. Financial supports from DFG and 
HSP III are gratefully acknowledged.

\section*{References}

1. Achucarro, A., Ortiz, M. (1993).
Phys. Rev. {\bf D 48}, 3600.

\noindent 2. Kiem, Y., Park, D. (1996).
Phys. Rev. {\bf D 53}, 747.

\noindent 3. Park, D.,  Kiem, Y. (1996).
Phys. Rev. {\bf D 53}, 5513.

\noindent 4. Schmidt, H.-J. (1998). Int. J. Mod. Phys. 
{\bf D 7},  215; gr-qc/9712034. For shorter versions see: 
DPG-Tagung 1998 Regensburg, S. 1029, Verh. Nr. 5 and Abstract
Conference GR 15 on General Relativity Poona/Indien
 1997, p. 214.

\noindent 5. Moessner, R., Trodden, M. (1995).
Phys. Rev. {\bf D 51}, 2801.

\noindent 6. Gergely, L. (1999).
A spherically symmetric closed universe as an example
of a $2D$ dilatonic model
gr-qc/9902016; Phys. Rev. D in print.

\noindent 7.   Eliezer, D. (1989).  Nucl. Phys. {\bf B 319},
667.

\noindent 8.  Organi, M. (1997). Class. Quant. Grav. {\bf 14},
1079.


\noindent 9.  Schmidt, H.-J. (1991). J. Math. Phys. {\bf 32},
1562.

\noindent 10.  Schmidt, H.-J. (1992).
 p. 330 in:  Relativistic Astrophysics and Cosmology,    
Eds.: S. Gottl\"ober, J.  M\"ucket, V. M\"uller, 
     WSPC Singapore.

\noindent 11. Solodukhin, S. (1995).
Phys. Rev. {\bf D 51}, 591.

\noindent 12. Mignemi, S.,  Schmidt, H.-J. (1995). Class.
Quant. Grav.
{\bf 12}, 849.

\noindent 13.  Balbinot, R.,  Fabbri, A. (1996).
Class. Quant. Grav. {\bf 13}, 2457.

\noindent 14. Bernutat, C. (1996). Thesis, University Potsdam,
86 pages.

\noindent 15.  Callan, C., Giddings, S., Harvey, J.,
Strominger, A.
(1992). Phys. Rev.  {\bf D 45}, 1005.

\noindent 16.  Cruz, J., Navarro-Salas, J., Navarro, M.,
Talavera, C.
(1997). Phys. Lett. {\bf B 402}, 270.

\noindent 17. Christensen, D.,  Mann, R. (1992). 
Class. Quant. Grav. {\bf 9}, 1769.

\noindent 18. Kl\"osch, T., Strobl, T. (1998). 
Phys. Rev. {\bf D 57}, 1034.

\noindent 19. Grosse, H., Kummer, W., Presnajder, P., 
Schwarz, D. (1992).
     J. Math. Phys. {\bf 33},  3892.

\noindent 20.  Buchbinder, I., Shapiro, I., Sibiryakov, A.
(1995).
Nucl. Phys. {\bf B 445}, 109.

\noindent 21. Kim, S., Lee, H. (1998). p. 332 in: Current
Topics in
Mathematical Cosmology, Eds.: Rainer, M., Schmidt, H.-J.,
WSPC Singapore.

\noindent 22. 
  Witten, E. (1991).
     Phys. Rev. {\bf D 44}, 314.

\noindent 23. 
 Frolov, V. (1992). Phys. Rev.  {\bf D 46}, 5383.

\noindent 24. 
  Mann, R. (1992).
Gen. Rel. Grav.   {\bf 24}, 433.

\noindent 25. 
  Myung, Y., Kim, J. (1996).
Phys. Rev. {\bf D 53}, 805.

\noindent 26. 
 Mignemi, S. (1994).
Phys. Rev. {\bf D 50}, R4733.


\noindent 27. 
 Azreg-Ainou, M. (1999).
Class. Quant. Grav. {\bf 16}, 245.

\noindent 28. 
 Chan, K., Mann, R. (1995).
Class. Quant. Grav. {\bf 12}, 1609.

\noindent 29. 
 Ahmed, M. (1996).
Phys. Rev. {\bf D 53}, 4403.

\noindent 30. 
   Mann, R.,  Ross, S. (1992). 
Class. Quant. Grav. {\bf 9}, 2335.

\noindent 31. 
 Peleg, Y., Bose, S., Parker, L. (1997). 
Phys. Rev. {\bf D 55}, R4525.

\noindent 32. 
  Kl\"osch, T., Strobl, T. (1996). Class. Quant. Grav.
{\bf 13}, 965, and 2395.

\noindent 33. 
Kl\"osch, T., Strobl, T. (1997). Class. Quant. Grav.
{\bf 14}, 1689.

\noindent 34. 
  Mann, R., Morris, M., Ross, S. (1993).
Class. Quant. Grav. {\bf 10}, 1477.

\noindent 35. 
 Chan, J., Mann, R. (1995).
Class. Quant. Grav. {\bf 12}, 351.

\noindent 36. 
  Cooperstock, F., Faraoni, V. (1995).
Gen. Rel. Grav. {\bf 27}, 555.

\noindent 37. 
 Mann, R., Sikkema, A. (1995). 
     Gen. Rel. Grav. {\bf 27},  563.

\noindent 38. 
 Fabbri, A. (1998).
Class. Quant. Grav. {\bf 15}, 373.

\noindent 39. 
 Kiem, Y., Park, D. (1996).
     Phys. Rev. {\bf D 53}, 747. 

\noindent 40. 
 Mielke, E., Gronwald, F., Obukhov, Yu., Tresguerres, R.,
Hehl, F. (1993). Phys. Rev. {\bf D 48}, 3648.

\noindent 41. 
 Yan, J., Qiu, X. (1998).
Gen. Rel. Grav. {\bf 30}, 1319.

\noindent 42. 
 Kummer, W., Schwarz, D. (1992).
Nucl. Phys. {\bf B 382}, 171.

\noindent 43. 
 Mignemi, S. (1996).
Mod. Phys. Lett. {\bf A 11}, 1235.

\noindent 44. 
 Katanaev, M., Kummer, W., Liebl, H. (1996).
Phys. Rev. {\bf D 53}, 5609.

\noindent 45. 
 Mignemi, S. (1997).
Ann. Phys. NY {\bf 257}, 1.

\noindent 46. 
 Katanaev, M. (1997).
J. Math. Phys. {\bf 38}, 946.

\noindent 47. 
 Ertl, M., Katanaev, M., Kummer, W. (1998).
Nucl. Phys. {\bf B 530}, 457.

\noindent 48. 
 Gegenberg, J., Kunstatter, G., Strobl, T. (1997).
Phys. Rev. {\bf D 55}, 7651.

\noindent 49. 
 Cadoni, M. (1996).
Phys. Rev. {\bf D 53}, 4413.

\noindent 50. 
  Fulling, S. (1986).
Gen. Rel. Grav. {\bf 18}, 609.

\noindent 51. 
 Amelino, G. et al. (1996).
Phys. Lett. {\bf B 371}, 41.

\noindent 52. 
  Teo, E. (1994).
Gen. Rel. Grav. {\bf 26},  13.

\noindent 53. 
  Fujiwara. T., et al. (1996).
Phys. Rev. {\bf D 53}, 852.

\noindent 54. 
  Page, D. (1996).
gr-qc/9603005; (1997). Class. Quant. Grav. {\bf 14}, 3041.

\noindent 55. 
  Solodukhin, S. (1996). 
Phys. Rev. {\bf D 53}, 824. 

\noindent 56. 
 Hawking, S. (1992).
     Phys. Rev. Lett. {\bf 69}, 406.

\noindent 57. 
  Vaz, C., Witten, L. (1995). 
Class. Quant. Grav. {\bf 12}, 2607.

\noindent 58. 
 Bose, S., Parker, L., Peleg, Y. (1996). 
Phys. Rev. Lett. {\bf 76}, 861.

\noindent 59. 
 Mikovic, A., Radovanovic, V. (1997).
Class. Quant. Grav. {\bf 14}, 2647.

\noindent 60. 
 Martinec, E. (1996).
Class. Quant. Grav. {\bf 13}, 1.

\noindent 61. 
  Deser, S. (1996).
 Found. Phys. {\bf 26}, 617.

\noindent 62. 
 Cadoni, M., Mignemi, S. (1999). 
Asymptotic symmetries of $AdS_2$
 and conformal group in $d=1$, 
hep-th/9902040.

\noindent 63. 
 Schmidt, H.-J. (1997). Grav. and Cosmol. 
{\bf 3}, 185; gr-qc/9709071.

\noindent 64. 
  Rosquist, K., Uggla, C. (1991). 
J. Math. Phys. {\bf 32},  3412.

\noindent 65. 
 Sanchez, M. (1997).
Trans. Am. Math. Soc. {\bf 349}, 1063.

\noindent 66. 
 Katanaev, M., Kummer, W., Liebl, H. (1997).
Nucl. Phys. {\bf B 486}, 353.

\noindent 67. 
Lemos, J., Sa, P. (1994). 
Phys. Rev. {\bf D 49}, 2897.

\noindent 68. 
 Schmidt, H.-J. (1996). Phys. Rev. {\bf D 54}, 7906;
(1998). p. 288  in: 
Current Topics in
Mathematical Cosmology, Eds.: Rainer, M., Schmidt, H.-J.,
WSPC Singapore.

\noindent 69. 
 Cruz, J., Navarro-Salas, J., Navarro, M. (1998).
Phys. Rev. {\bf D 58}, 087501.

\end{document}